\documentclass[preprint,aps,showpacs]{revtex4}
\usepackage{amsfonts,amsmath}

\begin{document}

\title{Direct Evidence of Negative Refraction at Media with Negative $\epsilon$ and $\mu$}
\author{Didier Felbacq}
\affiliation{GES UMR-CNRS 5650\\
Universit\'{e} de Montpellier II\\
CC074, Place E. Bataillon\\
34095 Montpellier Cedex 05, France}

\date{\today}

\begin{abstract}
We analyze wave scattering by an interface between a positive refractive
index medium and a Left-Handed Material. We show that the negative
refraction effect can be deduced directly from Maxwell equations, without
references to causality or losses.
\end{abstract}
\maketitle

There has recently been a huge interest in the physics of electromagnetic
waves propagation in media where the permeability and the permittivity in
harmonic regime have both a negative real part. There is no a priori
condition on the sign of the real part of these quantities, contrarily to
their imaginary part, where energy conservation implies a positive sign,
provided that the time dependence be chosen as $\exp \left( -i\omega
t\right) ,\omega >0$ \cite{landau}. Such materials have been considered by
Veselago \cite{veselago}, but his work has been a dead-letter until it was
recently suggested that they could be realized as effective media in the low
frequency limit of heterogeneous materials \cite
{pendry1,smith1,smith2,smith3}. Some spectacular applications have already
been proposed \cite{pendry3}, though the subject is a very polemical one 
\cite{williams,touffe,garcia,valanju}. It is not our point here to enter
these polemical issues, but to try to clarify a certain number of very basic
points in the scattering theory of such medium. In particular, we address
the crucial problem of the radiation conditions in such media. By a direct 
computation, involving only the
Maxwell system in its distributational form, we compute the behavior of a
beam illuminating a homogeneous slab of left-handed materials, and show the
asymptotic form of the transmitted beam.

Let us consider the basic scattering problem of a plane wave illuminating a
diopter separating the vacuum $\left( \varepsilon _{0},\mu _{0}\right) $
from a left-handed material with relative parameters $\left( \varepsilon
,\mu \right) $. The only difficulty here is the problem of the outgoing wave
condition inside the left-handed medium. Indeed, let us assume $s$-polarized
waves and a $z$ independent plane wave, with wavenumber $k_{0}$,
illuminating the diopter under the incidence $\theta $. Then, denoting $%
\alpha =k_{0}\sin \theta $, the total electric field writes $u\left(
x\right) e^{i\alpha y}\mathbf{e}_{z}$ and satisfies: 
\begin{equation}
\frac{d}{dx}\left( \frac{1}{\mu }\frac{du}{dx}\right) +\left(
k_{0}^{2}\varepsilon -\frac{\alpha ^{2}}{\mu }\right) u=0  \label{eq1}
\end{equation}
in the Schwartz distributions meaning. Denoting 
\begin{equation}
\beta =\sqrt{k_{0}^{2}\varepsilon \mu -\alpha ^{2}},\beta =\sqrt{%
k_{0}^{2}-\alpha ^{2}},  \label{beta}
\end{equation}
we impose, for $x<0$, the radiation condition: $u=e^{i\beta x}+r\left(
\alpha ,k_{0}\right) e^{-i\beta x}$. But for $x>0$, the situation should be
discussed. In the original work by Veselago \cite{veselago}, this question
is solved in the following way: ''(...) it is seen immediately (...) that the
sign of $\mathbf{k}$ is reversed''. In \cite{pok} it is claimed that at the
interface between a regular and left-handed medium: ''negative refraction
follows immediately from the continuity of the tangential component of $k$
and normal component of [Poynting vector]'', this last statement is false
(it suffices to make the direct computation to see it), but in fact, the main
point at issue is the choice of a radiation condition, something that cannot
be deduced from Maxwell equations in the case of a diopter,
 contrarily to the transmission conditions.

First, we only have two possible choices for the transmitted wave: $t\left(
\alpha ,k_{0}\right) e^{i \alpha x} e^{-i\beta x}$ or $t\left( \alpha ,k_{0}\right)
e^{i \alpha x} e^{i\beta x}$ (indeed, we necessarily have the conservation of the
horizontal component of the wave vector).\ Usually, the second solution is
kept, as representing a plane wave moving from the interface. However, in a
left-handed material, the Poynting vector $P=\frac{1}{2\omega \mu }\mathbf{k}
$ is directed in the opposite sense to $\mathbf{k}$. It seems that it is
mainly on this basis that the choice is made to reverse the usual convention
and choose $t\left( \alpha ,k_{0}\right)e^{i \alpha x} e^{-i\beta x}$ as a transmitted
field, and then obtain a non usual transmission condition (negative index).
It is not immediate that this is the good choice, as waves with a negative
phase velocity are known to exist.

One possible justification, which we recall briefly \cite{ziol}, is
mathematical: we have to choose a branch of the square root to define
function $\beta $. In case of usual materials, and for a time dependence of $%
\exp \left( -i\omega t\right) $, the chosen branch is determined by $\sqrt{1}%
=1$ and a cut-line along $i\Bbb{R}^{-}$ for instance. Here, we have to
consider $\left( \varepsilon ,\mu \right) \rightarrow \beta \left(
\varepsilon ,\mu \right) $ as a function of two complex variables. In order
to define this function one possibility is to use the above defined square
root and write function $\beta $ as the composition of two functions: 
\[
\left( \varepsilon ,\mu \right) \rightarrow \varepsilon \mu -\alpha ^{2}%
\stackrel{\sqrt{}}{\rightarrow }\beta \left( \varepsilon ,\mu \right) 
\]
this definition is not the only one for defining function $\beta $. However,
it has the advantage of using one sole square root for every problem. Let us
start from the point $\left( 1,1\right) $ where we know that $\beta \left(
1,1\right) =\cos \theta $. If the couple $\left( \varepsilon ,\mu \right) $
then describes half a circle in the upper complex plane, then the phase of
the product $\varepsilon \mu $ describes the entire interval $\left[ 0,2\pi %
\right] $ so that $z=\varepsilon \mu $ describes a circle and thus crosses
the cut-line, so that afterwards this crossing, the image point $\left(
\varepsilon \mu ,\beta \left( \varepsilon ,\mu \right) \right) $ moves on \
the second sheet of the Riemann surface of $z\rightarrow z^{2}$ so of course
when $\left( \varepsilon ,\mu \right) =\left( -1,-1\right) $ we have by
continuity that $\beta \left( -1,-1\right) =-\beta \left( 1,1\right) $. This
is precisely this situation that one wants to avoid in the theory of
multi-defined functions, and hence the introduction of a cut-line. However,
it is not desirable, from a mathematical point of view, to write $\sqrt{%
1\times 1}=1$ and $\sqrt{\left( -1\right) \times \left( -1\right) }=-1$
because it means that we do not use a properly defined function. So a
possible way out is to assume that there are some losses in the LHM:
we write $\mu \left( \omega \right) \simeq -1$, and $\varepsilon \left(
\omega \right) =-1+i\eta \left( \omega \right) $. Then $\beta \simeq \pm
\left( 1-\frac{i\eta }{2}\right) $, and for preventing the exponential
growth of waves, we have to choose the $-$ determination. This can be
realized in a perfectly coherent way by defining the square root with a
cut-line on $\Bbb{R}^{+}$, $\sqrt{i}=\left( 1+i\right) \frac{\sqrt{2}}{2}$,
the function being defined by upper continuity on $\Bbb{R}^{+}$ (i.e. $\sqrt{%
x}=\sqrt{x+i0^{+}}$).

Direct numerical simulations have given evidences for negative refraction 
\cite{ziol,souk}. Here, we propose a justification given directly by Maxwell
equations, which does not require the use of losses or considerations on
branches of the square root. We start with a situation where there is no
possible ambiguity: we consider a homogeneous slab of width $h$ illuminated
by a plane wave. The slab is filled with a LHM $\left( \varepsilon
,\mu \right) $ and it is surrounded by the vacuum. This situation is very
simple, for there is no problem of outgoing wave conditions in that case
other than in vacuum. The Maxwell system in the distributions meaning allows
to obtain, in the same conditions as (\ref{eq1}, \ref{beta}): 
\[
\begin{array}{ll}
x<0 & :u=e^{i\beta _{0}x}+re^{-i\beta _{0}x} \\ 
0<x<h & :u=Ae^{i\beta x}+Be^{-i\beta x} \\ 
x>h & :u=te^{i\beta _{0}x}
\end{array}
\]
where: 
\[
\begin{array}{l}
A=\frac{1}{2}\left( 1+\mu \frac{\beta _{0}}{\beta }\right) te^{-i\beta h},B=%
\frac{1}{2}\left( 1-\mu \frac{\beta _{0}}{\beta }\right) te^{i\beta h}, \\ 
t=\dfrac{\left( \kappa ^{2}-1\right) e^{i\beta h}}{\kappa ^{2}-e^{2i\beta h}}%
,r=\dfrac{\kappa \left( e^{2i\beta h}-1\right) }{\kappa ^{2}-e^{2i\beta h}}
\end{array}
\]
and 
\[
\kappa =\frac{\beta +\mu \beta _{0}}{\beta -\mu \beta _{0}} 
\]
We note that $\beta$ is real because both $\epsilon$ and $\mu$ are negative.
Let us assume now that it is a monochromatic wavepacket that illuminates the
slab, with some spectral amplitude $\mathcal{A}\left( \alpha \right) $. Then
we get, as a transmitted field: 
\[
u_{t}=\int \mathcal{A}\left( \alpha \right) t\left( k_{0},\alpha \right)
e^{i\left( \alpha y+\beta _{0}x\right) }d\alpha . 
\]
The important point now is to note that for $\epsilon$ and $\mu <0$, $\left|
\kappa \right| <1$ contrarily to what happens for $\mu$ and $\epsilon >0$.
Therefore we get the celebrated series representing multiscattering inside
the slab \cite{pendry1}: 
\[
t=\left( \kappa ^{2}-1\right) \frac{-e^{-i\beta h}}{1-\kappa ^{2}e^{-2i\beta
h}}=\left( 1-\kappa ^{2}\right) \left( e^{-i\beta h}\right) \sum_{p}\kappa
^{2p}e^{-2ip\beta h} 
\]
\textbf{with a minus sign in the exponential terms}.

Note that this series and its physical interpretation are direct
consequences of Maxwell equations.

We recover the usual result that the transmitted field is a collection of
rays, whose first one is given by: 
\[
u_{t,0}\left( 0,y\right) =\int \mathcal{A}\left( \alpha \right) \left(
1-\kappa ^{2}\right) e^{i\left( \alpha y-\beta h\right) }d\alpha . 
\]
However the minus sign in front of $\beta $ implies that the barycenter of
the beam is displaced towards the left and not towards the right as usual (with 
the grating convention that $\theta$ is positive in the direct orientation):
there is a negative beam refraction.

Inside the slab, the same expansion can be obtained for $A$ and $B$, and the
first transmitted and reflected beams are: 
\begin{eqnarray*}
u^{+}\left( x,y;h\right) &=&\int \mathcal{A}\left( \alpha \right) \frac{1}{2}%
\left( 1+\mu \frac{\beta _{0}}{\beta }\right) \left( 1-\kappa ^{2}\right)
e^{i\left( \alpha y-2\beta h\right) }e^{i\beta x}d\alpha \\
u^{-}\left( x,y\right) &=&\int \mathcal{A}\left( \alpha \right) \frac{1}{2}%
\left( 1-\mu \frac{\beta _{0}}{\beta }\right) \left( 1-\kappa ^{2}\right)
e^{i\alpha y}e^{-i\beta x}d\alpha .
\end{eqnarray*}
Now we let tend $h$ to infinity in order to recover the behavior of the
diopter. From Riemann-Lebesgue lemma, or else from weak convergence of $%
\left( e^{i\left( 2\beta h\right) }\right) $ towards $0$, we get: 
\[
\lim_{h\rightarrow +\infty }u^{+}\left( x,y;h\right) =0 
\]
and the same result holds for higher order rays (both up and down).
At the diopter limit $h=+\infty$, only $u^{-}\left( x,y\right) $ exists in the
LHM. Now, we see that the transmitted beam inside the LHM is a sum of
plane waves $\exp i\left( \alpha y-\beta x\right) $, and consequently we
have to choose the minus sign convention, so that \textbf{we indeed have to
change the usual radiation condition}: the wavevector $\mathbf{k}=\left(
\alpha ,-\beta \right) $ is in a sens opposite to that of the Poynting
vector.

We have shown that provided that an isotropic homogeneous left-handed
material can exist (at least for a certain frequency),
 then the Maxwell system leads naturally to the choice of
a modified radiation condition inside this medium and to negative refraction.

\end{document}